\documentclass[smallextended]{svjour3}       
\usepackage{graphicx} 

\usepackage{hyperref}
\usepackage{epstopdf}

\usepackage{xparse}
\NewDocumentCommand{\ceil}{s O{} m}{%
  \IfBooleanTF{#1} 
    {\left\lceil#3\right\rceil} 
    {#2\lceil#3#2\rceil} 
}
\NewDocumentCommand{\floor}{s O{} m}{%
  \IfBooleanTF{#1} 
    {\left\lfloor#3\right\rfloor} 
    {#2\lfloor#3#2\rfloor} 
}

\usepackage{mathptmx}      
\usepackage{latexsym}
\begin{document}

\title{Analogue algorithm for parallel factorization of an exponential number of large integers}

\subtitle{I. Theoretical description}


\author{Vincenzo Tamma
}


\institute{
              Institut f\"{u}r Quantenphysik and Center for Integrated Quantum Science and Technology (IQ\textsuperscript{ST}), Universit\"{a}t Ulm, Albert-Einstein-Allee 11, D-89081 Ulm, Germany \\
              Tel.: +49 (731) 50-22781\\
              Fax: +49 (731) 50-23086\\
              \email{vincenzo.tamma@uni-ulm.de
}           
}

\date{Received: date / Accepted: date}

\maketitle

\begin{abstract}
We describe a novel analogue algorithm that allows the simultaneous factorization of an exponential number of large integers with  a polynomial number of experimental runs. It is the interference-induced periodicity of ``factoring'' interferograms measured at the output of an analogue computer that allows the selection of the factors of each integer  \cite{thesis,jmp,prl,foundations}. At the present stage the algorithm manifests an exponential scaling which may be overcome by an  extension of this method to correlated qubits emerging from n-order quantum correlations measurements. We describe the conditions for a generic physical system to compute such an  analogue algorithm. A particular example given by an ``optical computer'' based on optical interference will be addressed in the second paper of this series \cite{pra2}.
\end{abstract}

\section{Fundamental principle of our work}\label{idea}

Multiplying numbers  is much easier than the inverse problem of finding the factors of large integers. Indeed, the security of codes relies on the current inability of a fast solution of this problem but is endangered by the well-known Shor's factoring algorithm \cite{shor,shorexp,Lomonaco2002,nielsen,mermin}.

%

Recently, important works about the use of different kinds of exponential sums \cite{primenumbers} for factorization purposes have been published \cite{clauser,summhammer,gauss1,gauss2,gauss3,mack,merkel1,stefanak1,stefanak2,wolk,rangelov}. Our method differs from the past experimental realizations \cite{mehring,mahesh,peng,bigourd,weber,gilowsky,sadgrove} in three important simultaneous achievements \cite{thesis,jmp,prl}. First, the division of $N$  by the test factors $\ell$ is not pre-calculated, but it is performed by the experiment itself. Second, several test factors are tested simultaneously.  Third, a scaling property inherent in the recorded interferograms allows us to obtain the factors of an exponential number of large integers.

The core of Shor's factoring algorithm  stands on the measurement of the periodicity  in the dominant maxima of the quantum probability distribution at the output of a quantum computer \cite{mermin}.  In line with Shor's idea, the key behind the algorithm we present in this paper stands in the measurement of the periodicity in the maxima of Continuous Truncated Exponential Sums (CTES) by performing first-order ``factoring'' interference processes with a physical system. Interestingly, the number n of necessary experimental runs scales logarithmically with respect to the largest integer to be factored. A noteworthy theoretical result at the core of the algorithm is that the periodicity of the resulting n interference patterns as a function of a continuous physical parameter in a given range, when appropriately scaled, allows us to achieve  factorization of large numbers by simply looking at the interference maxima at integer values.

The paper is organized in the following way. In section II will be given some mathematical background about CTES in connection with the hyperbolic function. In section III we will describe the factoring algorithm for a generic physical system and the conditions this system must satisfy. We will provide a generalization of the described algorithm in Section IV. Section V and VI will address final remarks and perspectives of extensions to polynomial scaling methods of factorization, respectively.

\section{Hyperbolic function, CTES periodicity and factorization}

We define a continuous truncated exponential sum (CTES) in the form \cite{thesis,jmp,prl,foundations}
\begin{eqnarray}\label{cespattern}
{\cal I}^{(M,j)}(\xi)\equiv |s^{(M,j)}(f(\xi))|^{2},
\end{eqnarray}
where  $s^{(M,j)}$ is the modulo-squared  value of the generalized curlicue function \cite{berry,thesis,foundations}
\begin{eqnarray}\label{curlicue}
s^{(M,j)}(\zeta)\equiv \bigg\vert\frac{1}{M}\sum_{m=1}^{M} \exp\left[
2\pi  i (m-1)^j \zeta \right]\bigg\vert^2
\end{eqnarray}
of integer order $j\geq 1$, with $M\geq2$ interfering terms, and
\begin{eqnarray}\label{f(xi)}
f(\xi)\equiv\frac{1}{\xi}
\end{eqnarray}
is the hyperbolic function with continuous variable $0\leq\xi\leq 1$.

\begin{figure*}[h]
   \begin{center}
   \begin{tabular}{c}
 \includegraphics[width=1\textwidth,natwidth=610,natheight=642]{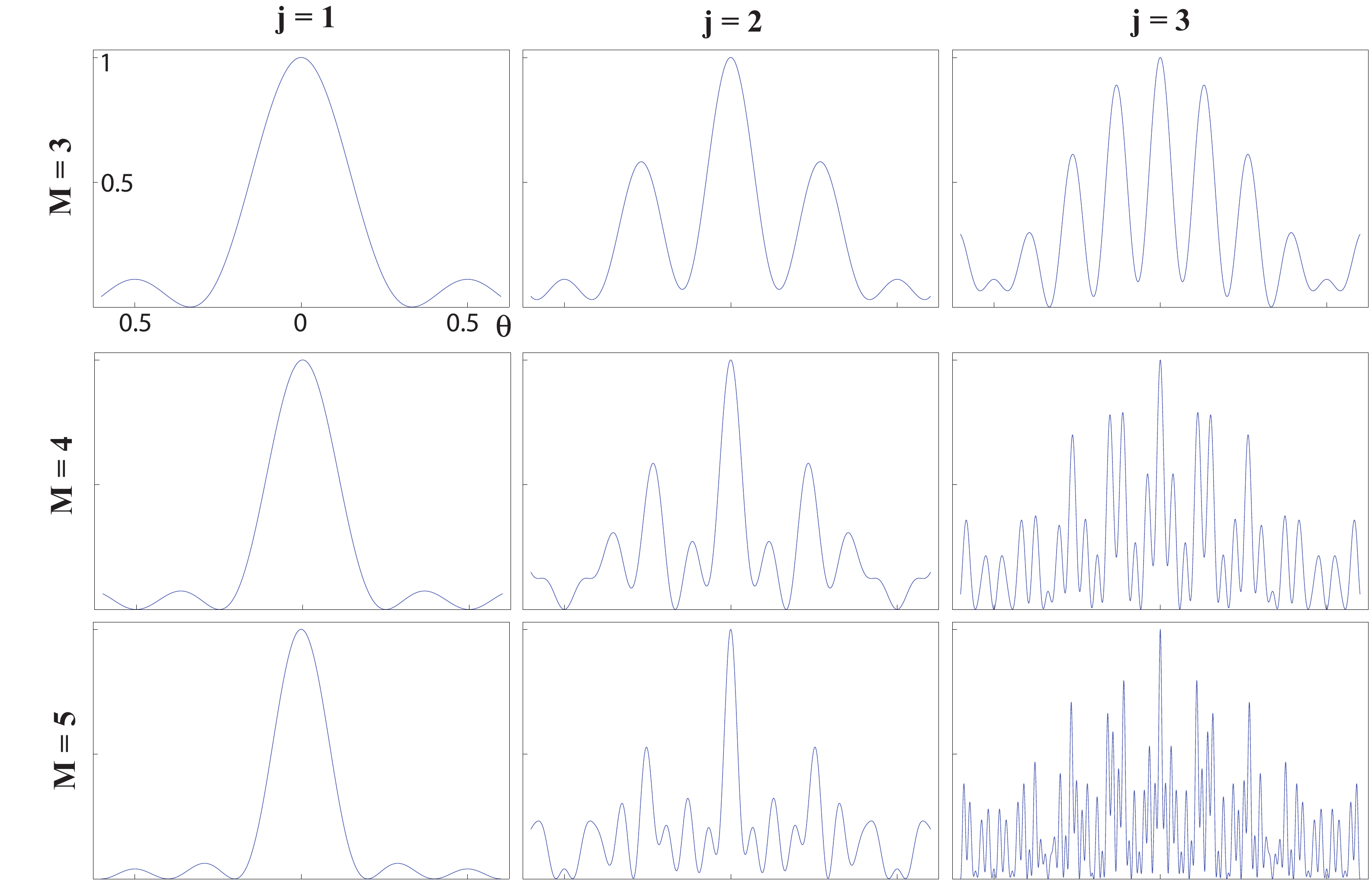}
   \end{tabular}
   \end{center}
\begin{quote}\caption [Modulo-squared value of the generalized curlicue function $s^{(M,j)}$  for $M=3,4,5$ and $j=1,2,3$.]{Modulo-squared value  $s_M^{(j)}$ of the generalized curlicue function in its dependence on the argument $\zeta$ for the number $M=3,4,5$ of interfering waves (column) and the power  $j=1,2,3$ of the phase shift (row). For increasing $M$ the dominant peak becomes narrower, which will make it easier in our algorithm to check if a peak corresponds to a factor or not.
 Unfortunately, at the same time the number of side maxima increases as well. However, we note that for a fixed $j$ and increasing $M$ the value of the maxima of second order decreases and $s_M^{(j)}$ becomes sharper. On the other hand,
  for a fixed $M$ but increasing $j$ the value of the maxima of second order increases and  $s_M^{(j)}$
  becomes wider \cite{foundations}.}\label{curlicuegraph}
\end{quote}\end{figure*}

In Fig. \ref{curlicuegraph}, we represent the modulo squared of the curlicue function $s^{(M,j)}=s^{(M,j)}(\zeta)$ in its dependence on the argument $\zeta$  for $M=3,4,5$ and $j=1,2,3$.
We note that $|s^{(M,j)}|^2$ has, for all the orders $j$, a dominant maximum at $\zeta=0$ with $s^{(M,j)}(0)=1$ and decaying oscillations on the sides. The higher the order $j$ and the truncation parameter $M$ of the curlicue function are, the larger is the number of decaying oscillations on the sides and the sharper is the dominant peak. Moreover, we recognize from Eq. (\ref{curlicue}) the periodicity property $s^{(M,j)}(\zeta + 1)=s^{(M,j)}(\zeta)$. Therefore, it suffices to consider $s^{(M,j)}=s^{(M,j)}(\zeta)$ in the domain $-1/2 \leq \zeta \leq 1/2$. In addition, $|s^{(M,j)}(\zeta)|^2$ is symmetric with respect to $\zeta=0$.

The hyperbolic function $f$  in Eq. (\ref{f(xi)}) induces in the function ${\cal I}^{(M,j)}$ a notable periodicity. Indeed, the function ${\cal I}^{(M,j)}$ is characterized by dominant maxima, which repeat each time $f$ assumes integer values.

Why does such a CTES periodicity matter in factorization?

In order to answer this question we first point out that, as shown in Refs. \cite{thesis,prl,foundations}, the factorization problem could be, in principle, solved if we can achieve the complete knowledge of the hyperbolic function $f$.

Indeed, if we look at $f$ as a function of the new variable $\xi_N$ obtained by the scaling relation
\begin{eqnarray}\label{scaling}
\xi_N \equiv N \xi,
\end{eqnarray}
we obtain
\begin{eqnarray}\label{f(xi)rescaled}
f(\xi_N)=\frac{N}{\xi_N}.
\end{eqnarray}
For each possible value of $N$, the factors are given by the integer values $\xi_N = \ell$ such that
\begin{eqnarray}\label{factors}
f(\ell)=\frac{N}{\ell}=k,
\end{eqnarray}
with $k$ a positive integer.

Unfortunately, it is not an easy task to compute the hyperbolic function so that for any given integer $N$ the condition (\ref{factors}) can be verified in order to identify the factors.

Interestingly, we can exploit the constructive/destructive periodic interference characterizing  the CTES function in Eq. (\ref{cespattern}) as a tool in the distinction between factors and non factors of any given number $N$. In particular, the  rescaled hyperbolic function $f(\xi_N)$ in Eq. (\ref{f(xi)rescaled}) corresponds to the rescaled CTES
\begin{eqnarray}\label{rescaledcespattern}
{\cal{I}}^{(M,j)}(\xi_N) = \bigg|\frac{1}{M}\sum_{m=1}^{M} \exp\left[
2\pi  i (m-1)^j f(\xi_N) \right]\bigg|^{2}
\end{eqnarray}
as a function of $\xi_N\equiv N\xi$.
The condition (\ref{factors}) leads to total constructive interference in the rescaled CTES function ${\cal{I}}^{(M,j)}(\xi_N)$ in Eq. (\ref{rescaledcespattern}). Indeed, the factors of an arbitrary number $N$ are the integer values $\ell$ of $\xi_N$ corresponding to  dominant maxima in the function ${\cal{I}}^{(M,j)}$.

\begin{figure*}[t]
   \begin{center}
   \begin{tabular}{c}
\includegraphics[width=1\textwidth,natwidth=610,natheight=642]{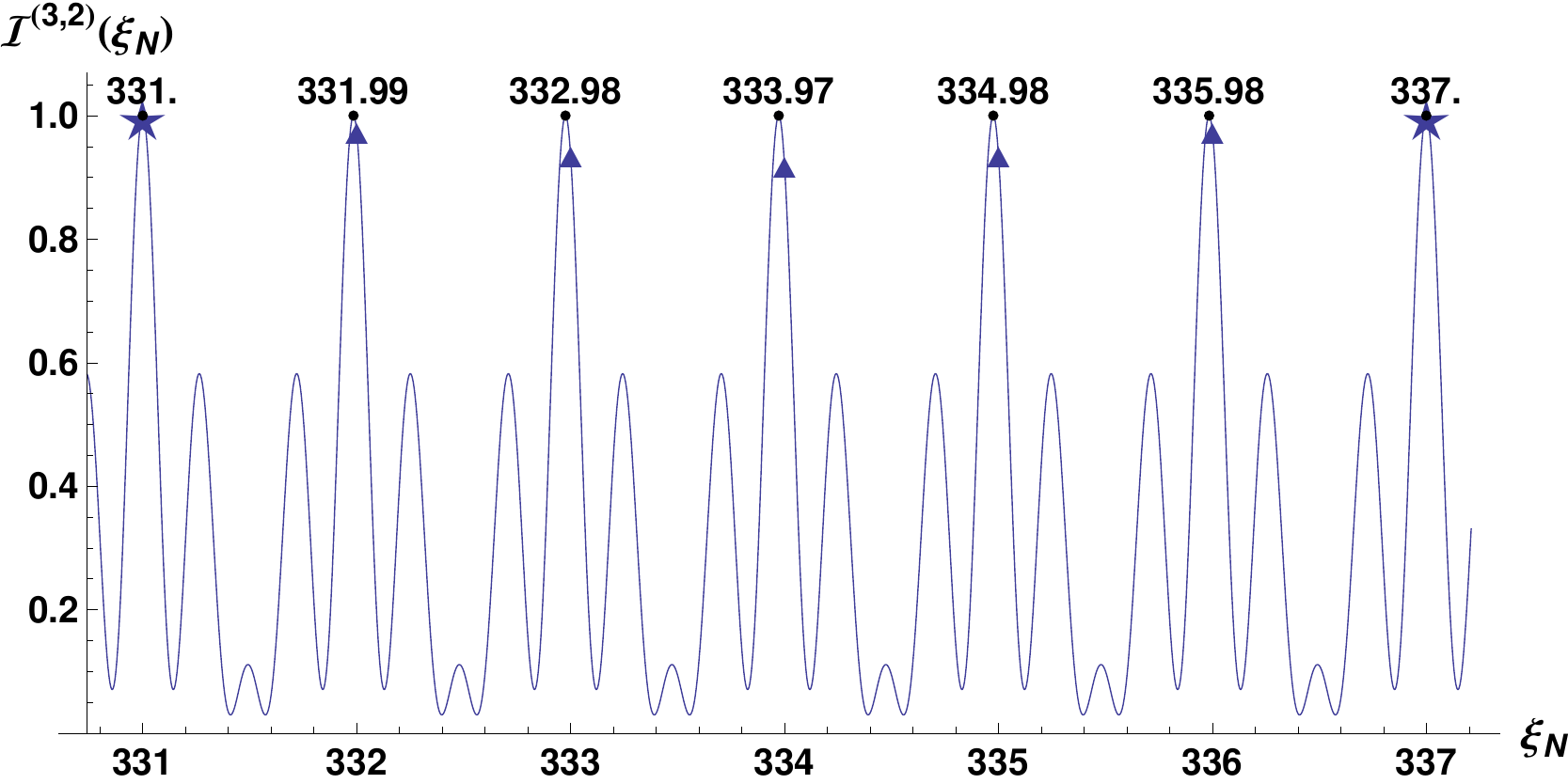}
  \end{tabular}
   \end{center}
\begin{quote}\caption{Rescaled CTES (j=2) function ${\cal{I}}^{(M,j)}(\xi_N)$ in Eq. (\ref{rescaledcespattern})  with $M=3$ for $N=111547$, as a function of the variable $\xi_N\equiv N \xi$ in the interval $[330.84,337.21]$  \cite{jmp}. We can see that the two factors $\xi_N=331,337$, represented by stars, correspond to  complete constructive interference, with respect to the other integer trial factors $\xi_N=332,333,334,335,336$, represented by triangles, which present partially destructive interference.\label{contexp}}
\end{quote}
\end{figure*}

In Fig. \ref{contexp} is simulated the rescaled CTES function in Eq. (\ref{rescaledcespattern}), with $M=3$, $j=2$, as a function of the variable $\xi_N \in [330.84,337.21]$ for the factorization of $N=111547$. We can see that the two factors $\ell=331,337$ (represented by stars) correspond to complete constructive interference. On the other hand, for the other test factors (represented by triangles) there is partially destructive interference. Moreover, there are absolute maxima (represented by points) which do not correspond to integer test factors.

\begin{figure*}[h]
   \begin{center}
   \begin{tabular}{c}
\includegraphics[width=1\textwidth,natwidth=610,natheight=642]{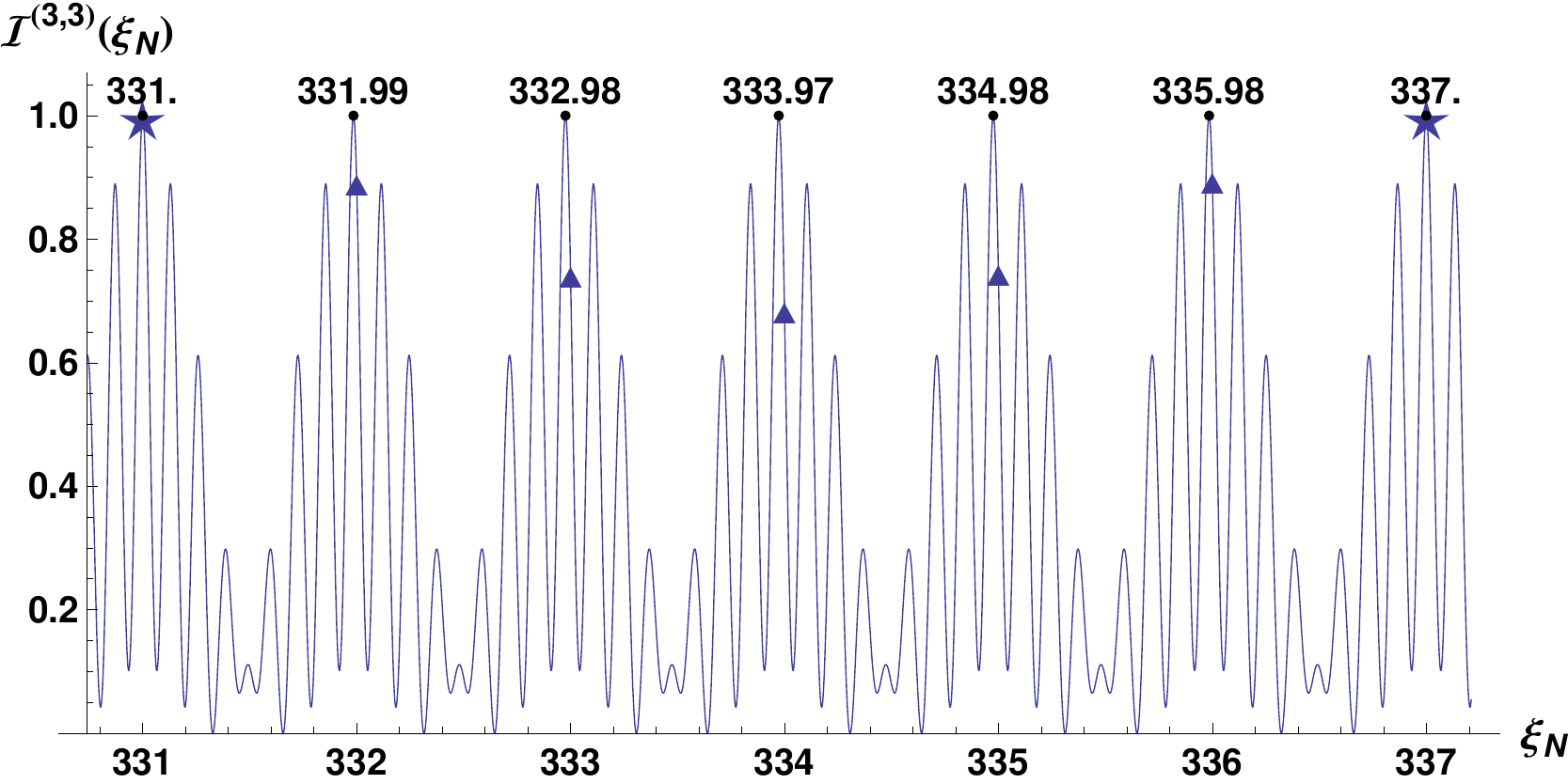}
  \end{tabular}
   \end{center}
\begin{quote}\caption{Rescaled CTES (j=3) function ${\cal{I}}^{(M,j)}(\xi_N)$ in Eq. (\ref{rescaledcespattern})  with $M=3$ for $N=111547$, as a function of the variable $\xi_N\equiv N \xi$ in the interval $[330.84,337.21]$ \cite{jmp}. The two factors $\xi_N = 331, 337$, represented by
stars, correspond to complete constructive interference, with respect to the other integer trial factors $\xi_N=332,333,334,335,336$ in such an interval,
represented by triangles. As expected, the peaks associated with the absolute maxima in the case $j=3$ are sharper than the respective peaks in the case $j=2$ represented in Fig. \ref{contexp}. On the other hand, the value of the maxima of second order in the function ${\cal{I}}^{(M,j)}$ increases at the increasing of the order $j$. \label{contexp2}}
\end{quote}
\end{figure*}

In Fig. \ref{contexp2}, instead, we have simulated  the CTES function in the case of $j=3$ 
 for the same value of $N$ and $M$ and the same range of values of $\xi_N$.
As expected, it turns out that as the order $j$ of the exponential sum increases, the peaks associated with the absolute maxima become sharper. On the other hand,  the values of the second order maxima in the interference pattern increase for larger orders $j$. In order to suppress such maxima, it is necessary to increase the number of terms $M$ in the sum. Also, in such a case, the first order peaks become increasingly sharper.

If we consider the range of values of $\xi_N$ in Fig. \ref{contexp} and Fig. \ref{contexp2}, the closer a non factor is to a factor, the larger the corresponding value is of the rescaled CTES function.
Of course, if we move to integer values of $\xi_N$ farther from the factors there is a larger probability of partial or total destructive interference between the terms in the CTES function. This makes the distinction between factors and non factors easier.

\begin{figure*}[t]
   \begin{center}
   \begin{tabular}{c}
   \includegraphics[width=1\textwidth,natwidth=610,natheight=642]{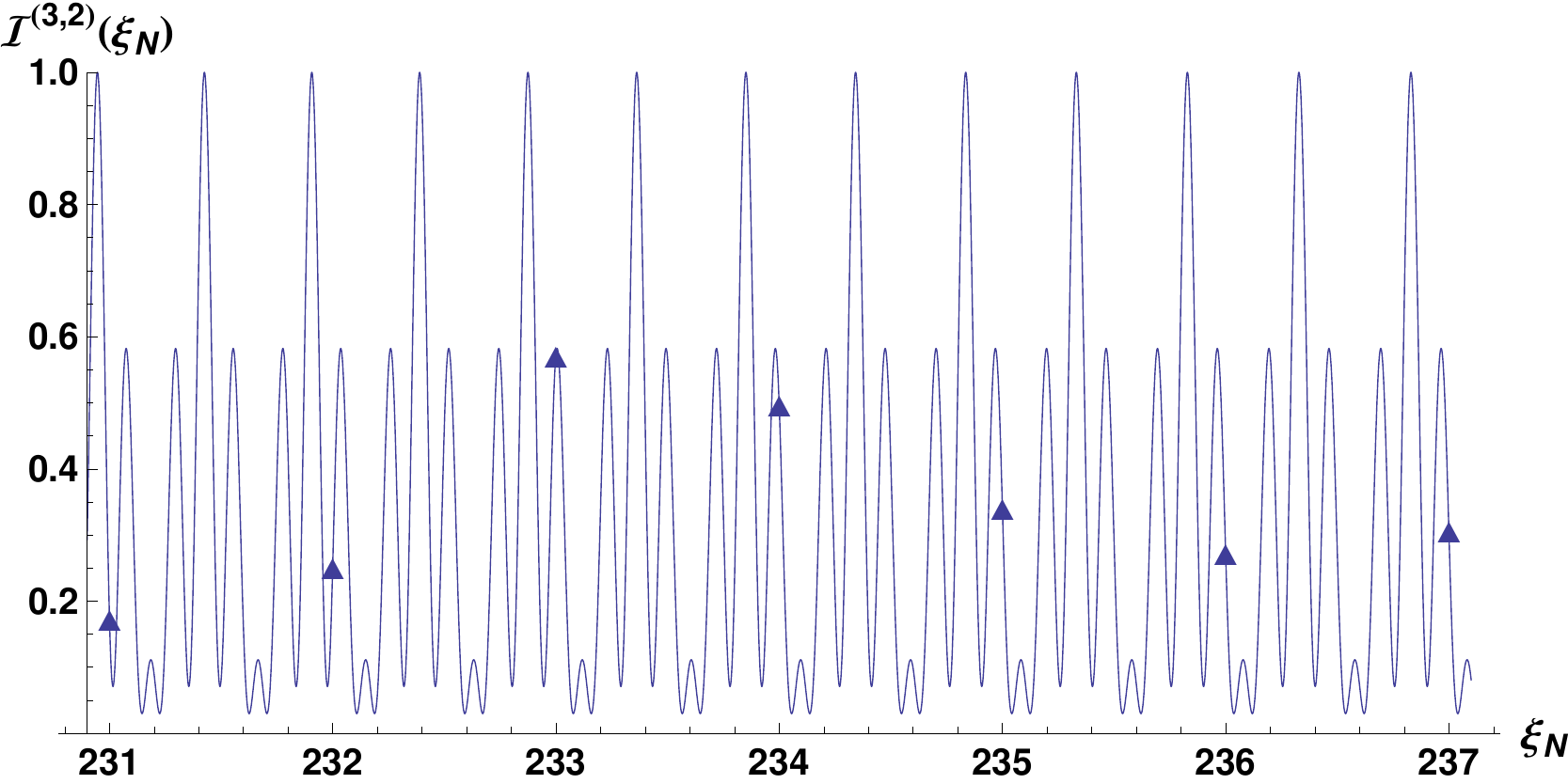}
   \end{tabular}
   \end{center}
\begin{quote}\caption [Simulation of the rescaled CTGS function for test factors far from the factors]{Rescaled CTES ($j=2$) function ${\cal{I}}^{(M,j)}(\xi_N)$ in Eq. (\ref{rescaledcespattern})  with $M=3$ for $N=111547$, as a function of the variable $\xi_N \equiv N \xi$ in the interval $[230.9, 237.1]$. We can clearly see that all the integer trial factors $\xi_N=231,232,233,234,235,236,237$ in such a range, represented by triangles, correspond to a  relatively small value of the CTES function with respect to the dominant  maxima so they can be easily disregarded as possible factors.\label{simulation}}
\end{quote}\end{figure*}

In general,  the maximum possible step $\Delta \xi$ between consecutive values of $\xi$ where the CTES has to be computed can not be exactly determined since it depends on the value of the factors of the integer $N\leq N_AA{max}$ to be factored. However, it can be easily shown \cite{thesis} that such value is included in  the interval
\begin{eqnarray}\label{deltaxiN1}
N^{-2} < \Delta \xi < 1  
\end{eqnarray}
 if $\xi_N \in [1, \sqrt{N}]$, or the interval 
 \begin{eqnarray}\label{deltaxiN2}
 \sqrt{N^{-3}} < \Delta \xi < 1
 \end{eqnarray}
 if $\xi_N \in [\sqrt{N}, N]$.

In conclusion, the CTES  function in Eq. (\ref{cespattern}) allows us to extract the information about factors encoded by its periodicity in the dominant maxima. Such a periodicity is imprinted by its functional dependence on the hyperbolic function $f(\xi) \equiv 1/\xi$. We recognize all the values $\xi$ corresponding  to an integer value of $f(\xi)\equiv1/\xi$ as dominant maxima in the interference pattern. When for one of these values of $\xi$ we find, for a given large number $N$, $\xi_N\equiv N\xi$ be an integer, such an integer is a factor of $N$.

However, a digital implementation of the CTES function for factoring purposes would rely on performing an exponential number of divisions, which are expensive operations for digital computers. On the other hand, an analogue implementation  of such a ``factoring'' function would rely on the help of ``nature'' to perform such divisions for us. In the next section, we will describe the prerequisites for a generic physical system to compute a factoring algorithm based on the implementation of CTES interferograms.

\section{Factoring analogue algorithm}

In the previous section we have shown how the implementation of the CTES function ${\cal{I}}^{(M,j)}(\xi)$ in Eq. (\ref{cespattern})  would allow, in principle, the prime number decomposition of arbitrary integers. Now, we want to describe the analogue implementation of the CTES algorithm with a generic physical system that exploits interference. This will lead us to the introduction of a two-dimensional interferogram $I^{(M,j)}(o_{\xi};x)\equiv {\cal{I}}^{(M,j)}(\xi)$ as a function of a continuous physical parameter $o_{\xi}\equiv\xi x$ and a discrete physical parameter $x$ associated with two independent observables $O_{\xi}$ and $O_x$, respectively.
In the second paper of this series \cite{pra2} we will give an example of an ``optical computer" based on a multi-path Mach Zehnder interferometer, where the physical parameter $o_{\xi}$ and $x$ will correspond to the wavelengths associated with the spectrum of the optical source and the optical-path unit, respectively.   

We will first introduce a CTES ``factoring'' procedure which takes advantage of a single interferogram $I^{(M,j)}(o_{\xi};x)$ associated with a given value of $x$. We will determine the range $N_{min}\leq N\leq N_{max}$ of factorable numbers $N$ by covering all the trial factors  in either the range $[1,\sqrt{N}]$ or $[\sqrt{N},N]$  with a given range $o_{min}\leq o_{\xi} \leq o_{max}$ of values of the observable $O_{\xi}$  \cite{thesis}.

Next, we will show that the CTES ``factoring'' algorithm can exploit the degree of freedom in the variable $x$ of the analogue function $I^{(M,j)}(o_{\xi};x)$ in order to check all the necessary trial factors of an exponential number of large integers $N$  \cite{thesis}. In particular, such a procedure is based on the measurement of the periodicity of a number $n$ of interferograms for different suitable values of $x$. We will interestingly find that the value of $n$ scales logarithmically with respect to the largest number $N_{max}$ to be factored.

%

\subsection{Analogue implementation of the CTES function}

We have shown that the implementation of a CTES function in Eq. (\ref{cespattern}) would allow the factorization, in principle, of arbitrary numbers. Unfortunately, the calculation of such a sum would require an exponential number of divisions associated with the computation of the function $f(\xi)$.  On a digital computer, for which division is a rather costly process, such a computation turns out to be very slow. Therefore, it would be more efficient to reproduce the CTES function with an analogue technique in order to solve the problem quickly. More explicitly, an analogue implementation of the CTES algorithm is possible if there is a physical system able to compute divisions for us and then read out the factors by taking advantage of the interference process leading to CTES interferograms.

In particular such a physical system needs to fulfill three main requirements.

First, we assume that the system is characterized by two independent observables $O_x$ and $O_{\xi}$.
The observable $O_x$ can be tuned to values
\begin{eqnarray}\label{omxj}
o_{x}^{(m,j)} \equiv  (m-1)^j x,
\end{eqnarray}
with $j$ a positive integer and  $m=1,...,M$. The unit of measurement $x$ can be, in principle, arbitrarily varied.
On the other hand, the observable  $O_{\xi}$ assumes a continuous range of values
\begin{eqnarray}\label{oxi}
o_{\xi} \equiv  \xi x,
\end{eqnarray}
where $x$ is the unit of measurement chosen in Eq. (\ref{omxj}).

Second, the physical system needs to be able to compute the hyperbolic function in Eq. (\ref{f(xi)}) in the form of ratios between the values of $O_{x}$ and $O_{\xi}$ for a given interval of the variable $\xi$.

Third, the system must be able to exploit interference in order to reproduce the interferogram
\begin{eqnarray}\label{cespatternoxoxi}
I^{(M,j)}(o_{\xi};x)\equiv \bigg|\frac{1}{M} \sum_{m=1}^{M} \exp\left[
i\frac{o_{x}^{(m,j)}}{o_{\xi}}\right]\bigg|^{2} \equiv {\cal{I}}^{(M,j)}(\xi)\nonumber\\
\end{eqnarray}
as a function of  $o_{\xi}\equiv  \xi x$.
Of course, for a parallel evaluation of the sum for several values of $\xi$ it is necessary that the system contains at the same time the information about all the possible values of the physical observable $O_{\xi}$. The corresponding CTES function ${\cal{I}}^{(M,j)}(\xi)$   in Eq. (\ref{cespattern}) will finally allow us to extract the information about factors.

In particular, for a generic value of the number $N$ to be factored, it is possible to look at the obtained interferogram in Eq. (\ref{cespatternoxoxi}) as a function of the rescaled variable in Eq. (\ref{scaling})
\begin{eqnarray}\label{xiN}
\xi_N \equiv N \xi = \frac{N}{x}o_{\xi},
\end{eqnarray}
where we use Eq. (\ref{oxi}) in the second equality.

Indeed, we obtain the rescaled interferogram
\begin{eqnarray}\label{rescaledanaloginterf}
I^{(M,j)}(\xi_N;x)\equiv {\cal{I}}^{(M,j)}(\xi_N)
\end{eqnarray}
corresponding to the rescaled CTES function ${\cal{I}}^{(M,j)}(\xi_N)$ defined in Eq. (\ref{rescaledcespattern}).

Each time there is a  dominant maximum at a value of $o_{\xi}$ for which $\xi_N$ in Eq. (\ref{xiN}) is an integer, we find such an integer to be a factor of $N$.

It is important to point out that the information about the one-dimensional CTES function ${\cal{I}}^{(M,j)}(\xi)$  in Eq. (\ref{cespattern}) is inferred by the two-dimensional CTES interferogram $I^{(M,j)}(o_{\xi};x)$ in Eq. (\ref{cespatternoxoxi}) as a function of the physical variable $o_{\xi}$ and  the physical parameter $x$.  Indeed, as becomes clear later, this feature will turn out to be one of the key points to understand the working principle behind the CTES analogue algorithm.

\subsection{Factorization with a single interferogram}
In this section, we describe the ``factoring'' procedure based on the use of a single  interferogram given by  the CTES analogue function $I^{(M,j)}(o_{\xi};x)$ in Eq. (\ref{cespatternoxoxi}) recorded at a given value of $x$. We address the question of the interval $N_{min,x}\leq N\leq N_{max,x}$ of numbers $N$  factorable by covering all the trial factors  in either the range $[3,\sqrt{N}]$  \footnote{We are sure that there is at least one factor of $N$ in such interval. The trial factor $2$ is obviously excluded since it is easy to recognize if $N$ is an even integer.} or $[\sqrt{N},N]$  with a given range $o_{min}\leq o_{\xi} \leq o_{max}$ of values for the observable $O_{\xi}$.

 In general, for each integer $N$ a generic trial factor $\ell$ can be checked only if 
\begin{eqnarray}\label{lminlmaxNo}
\xi_N = \ell \in [\xi_{N}^{(min)},\xi_{N}^{(max)}] \equiv [\frac{N}{x} o_{min}, \frac{N}{x} o_{max}],
\end{eqnarray}
where $\xi_{N}^{(min)}$ and $\xi_{N}^{(max)}$ are respectively the smallest and largest values that the variable $\xi_N$ in Eq. (\ref{xiN}) can assume for the rescaled interferogram $I^{(M,j)}(\xi_N;x)$.

We consider the case in which we want to check all the  trial factors $\ell \in[3,\sqrt{N}]$   leading  from Eq. (\ref{lminlmaxNo}) to the condition
\begin{eqnarray}\label{band1N}
\mbox{Method $(1)$:} \ \xi_N = \ell \in[3,\sqrt{N}]\subseteq [\frac{N}{x} o_{min}, \frac{N}{x} o_{max}].
\end{eqnarray}
By dividing the upper and lower bounds of each interval respectively by $\sqrt{N}$ and $3$ we find
\begin{eqnarray}\label{band1N2}
1 \in [\frac{N}{3 x} o_{min}, \frac{\sqrt{N}}{x} o_{max}],
\end{eqnarray}
which implies
 \begin{eqnarray}\label{band1N3}
  \frac{N}{3 x} o_{min} \leq 1 \leq \frac{\sqrt{N}}{x} o_{max}
 \end{eqnarray}
 for each integer $N$ in the interval $N_{min,x}\leq N\leq N_{max,x}$ to be determined. 
By squaring the third term in this series of inequalities we obtain the condition
\begin{eqnarray}\label{xlimit}
 x\leq x^{(1)}\equiv \displaystyle\frac{3 o_{max}^2}{o_{min}},
\end{eqnarray}
giving an upper bound to the choice of the parameter $x$ associated with the rescaled interferogram $I^{(M,j)}(\xi_N;x)$ independent of the number $N$ to factorize.
From Eq. (\ref{band1N3}) we also easily obtain \footnote{We recall that for any real number $y$ the ceiling $\ceil*{y}$ is the smallest integer larger than $x$, while the floor $\floor*{y}$ is the largest integer lower than $y$.}
\begin{eqnarray}
 & N_{max,x}^{(1)} \equiv \floor*{\displaystyle\frac{3 x}{o_{min}}} \nonumber\\ 
 & \mbox{and}\nonumber \\ 
& N_{min,x}^{(1)} \equiv \ceil*{\displaystyle\frac{x^2}{o_{max}^2}},
\end{eqnarray}
defining the interval of factorable numbers with a single interferogram associated with a given value $x$ in Eq. (\ref{xlimit})  in the range $o_{min}\leq o_{\xi} \leq o_{max}$.

In  the particular case of an interferogram recorded at the maximum possible value $x=x^{(1)}$ in Eq. (\ref{xlimit}), the condition (\ref{band1N3}) reads
$$\frac{N}{3 x^{(1)}} o_{min} = \frac{\sqrt{N}}{x^{(1)}} o_{max}= 1. $$
If $o_{max} / o_{min}$ is an integer, we find the largest but also the only integer
\begin{eqnarray}\label{Nmax1}
N^{(1)}\equiv \displaystyle  \frac{9 o_{max}^2}{o_{min}^2}
\end{eqnarray}
factorable by using a single experimental interferogram 
$I^{(M,j)}(o_{\xi};x)$ in Eq. (\ref{cespatternoxoxi}) recorded for the value $x = x^{1}$.

We consider now the second method in which we want to check all the  trial factors $\xi_N = \ell \in [\sqrt{N},N]$  leading from Eq. (\ref{lminlmaxNo}) to the condition
\begin{eqnarray}\label{band2N}
\mbox{Method $(2)$:} \ \xi_N = \ell \in[\sqrt{N}, N]\subseteq [\frac{N}{x} o_{min}, \frac{N}{x} o_{max}].
\end{eqnarray}
By dividing the lower bound and the upper bound of each interval by $\sqrt{N}$ and $N$, respectively, we find
\begin{eqnarray}\label{band2N2}
1 \in [\frac{\sqrt{N}}{x} o_{min}, \frac{o_{max}}{x} ],
\end{eqnarray}
which implies
\begin{eqnarray}\label{band2N3}
1 \geq\frac{\sqrt{N}}{x} o_{min} \leq \frac{o_{max}}{x} \geq 1
 \end{eqnarray}
for each integer $N$ in the interval $N_{min,x}\leq N\leq N_{max,x}$ to be determined. 
From the last inequality in Eq. (\ref{band2N3}) we obtain the condition 
\begin{eqnarray}\label{xlimit2}
 x\leq x^{(2)}\equiv o_{max}.
\end{eqnarray}
From Eq. (\ref{band2N3}) we also easily obtain 
\begin{eqnarray}
 & N_{max,x}^{(2)} \equiv \floor*{\displaystyle\frac{x^2}{o_{min}^2}} \nonumber\\ 
 & \mbox{and}\nonumber \\ 
& N_{min}^{(2)}\equiv 1,
\end{eqnarray}
defining the interval of factorable numbers with a single interferogram associated with the generic value $x$ in Eq. (\ref{xlimit2}) for the given range $o_{min}\leq o_{\xi} \leq o_{max}$.

We consider now the case of a single interferogram recorded at the maximum value $x=x^{(2)}\equiv o_{max}$ in Eq. (\ref{xlimit2}) leading to the largest interval
\begin{eqnarray}\label{Nmax2}
N_{min}^{(2)}\equiv 1 \leq N \leq N_{max}^{(2)}\equiv \displaystyle  \floor*{\displaystyle\frac{o_{max}^2}{o_{min}^2}}
\end{eqnarray}
of factorable integers.

We finally demonstrate that with a single interferogram it is possible to factorize a  number $$\Delta N \sim N_{max}\sim \displaystyle  \frac{o_{max}^2}{o_{min}^2}\sim 2^{n_{max}}$$ of integers exponential with respect to the number of binary digits $n_{max}$ associeted with $N_{max}$.
However, in general, the largest factorable integer $N_{max}$ is limited by the value $o_{max}/o_{min}$ associated with the physical system. For this reason, in the next section we will describe a factorization procedure which takes  advantage of several interferograms  $I^{(M,j)}(o_{\xi};x)$  in Eq. (\ref{cespatternoxoxi}) at different values $x$ in order to factor  numbers exploiting a limited fixed range of values $o_{\xi}$ of a physical observable $O_{\xi}$. In such a method, the maximum factorable number $N_{max}$ will depend only on the largest value achievable for the parameter $x$.

\subsection{Factorization with a sequence of interferograms}\label{sequencealgo}
So far we have restricted ourselves to a factorization method involving a single interferogram $I^{(M,j)}(o_{\xi};x)$  in Eq. (\ref{cespatternoxoxi}) defined at a fixed value of the parameter $x$. However, the remarkable scaling property $\xi_N\equiv N o_{\xi}/x$ characterizing the function  $I^{(M,j)}(o_{\xi};x)$ allows us to vary the physical values of both $o_{\xi}$ and  $x$. This implies that we can, in principle, change arbitrarily both the number $N$ we are looking at and the relative trial factors to check by simply varying these two physical ``knobs''.

In particular, we seek to address the question of the largest number  $N_{max}$ that can be factored by measuring several interferograms $I^{(M,j)}(o_{\xi};x)$ corresponding to different values of $x$ in a fixed domain $o_{min}\leq o_{\xi}\leq o_{max}$ of the variable $o_{\xi}$.

\subsubsection{Factorization of a single integer $N$}
We consider first the case of a single generic large integer $N$ to be factored.
The scaling property $\xi_N\equiv N o_{\xi}/x$, written as $N \equiv \xi_N x/o_{\xi}$, implies that, in principle, there is no limit to the largest possible value of $N$ if we can vary the unit $x$ up to arbitrary large values. Of course, in practice, the maximum achievable value of $x$ is limited by the particular physical system we are using.
We now demonstrate that the number $n$ of values of $x$, for which the ``factoring'' interferogram $I^{(M,j)}(o_{\xi};x)$  in Eq. (\ref{cespatternoxoxi}) must be recorded, scales logarithmically with respect to the value of $N$.

It should first be pointed out that a single interferogram registered at a given value $x$ allows us to check only the trial factors
\begin{eqnarray}\label{lminlmaxN}
\xi_N = \ell \in [\xi_{N}^{(min)},\xi_{N}^{(max)}] \equiv [\frac{N}{x} o_{min}, \frac{N}{x} o_{max}],
\end{eqnarray}
which, in general, for the fixed domain $o_{min}\leq o_{\xi}\leq o_{max}$ of values $o_{\xi}$, may correspond only to a subset of the total range $3\leq \ell \leq \sqrt{N}$ or $\sqrt{N}\leq \ell \leq N$ of values $\xi_N=\ell$ we would need to cover in order to factor a generic integer $N$. For this reason, we consider a suitable sequence of values  $x=x_i$, with $i=0,1,...,n-1$. Each interferogram registered at the value  $x=x_i$, with $i=0,1,...,n-1$, allows us from Eq. (\ref{lminlmaxN}) to cover all the trial factors 
\begin{eqnarray}\label{liN}
\xi_N = \ell \in [\xi_{N,i},\xi_{N,i+1}] \equiv [\frac{N}{x_{i}}o_{min},  \frac{N}{x_{i}}o_{max}],
\end{eqnarray}
with $i=0,1,...,n-1$, where
\begin{eqnarray}
\xi_{N,i+1}\equiv \frac{N}{x_{i}}o_{max}\equiv \frac{N}{x_{i+1}}o_{min}
\end{eqnarray}
satisfies the condition for consecutive intervals.

This implies that the sequence $x_i$, with $i=0,...,n-1$, associated with the $n$ interferograms is defined by the condition
\begin{eqnarray}\label{x_iN}
x_{i+1}\equiv\frac{x_i}{c}  < x_i,
\end{eqnarray}
with
\begin{eqnarray}\label{c2}
c\equiv \frac{\xi_{N,i+1}}{\xi_{N,i}}=\frac{o_{max}}{o_{min}}>1
\end{eqnarray}
and, from Eq. (\ref{liN}),
\begin{eqnarray}\label{x0}
\frac{x_{0}}{o_{min}}\equiv  \frac{N}{\xi_{N,0}}.
\end{eqnarray}
From Eq. (\ref{c2}) we obtain 
\begin{eqnarray}
\xi_{N,n} = c ^n \xi_{N,0},
\end{eqnarray}
leading to the number
\begin{eqnarray}\label{n}
n \equiv log_{c} \frac{\xi_{N,n}}{\xi_{N,0}}
\end{eqnarray}
of interferograms necessary to cover all the trial factors
\begin{eqnarray}\label{lNtotal}
\xi_N = \ell \in [\xi_{N,0},\xi_{N,n}] \equiv [\frac{No_{min}}{x_{0}} ,\frac{N o_{min}}{x_{0}} \ c^n].
\end{eqnarray}

We deduce that the value $x_0$ for the first interferogram as well as the total number $n$ of interferograms $I^{(M,j)}(o_{\xi};x)$ depend on the integer $N$ to be factored and on the associated interval $[\xi_{N,0},\xi_{N,n}]$ of trial factors to be checked. In particular, factorization can be achieved for a generic integer $N$ only if
\begin{eqnarray}\label{range1}
&\mbox{Method $(1)$:} \\ &\xi_N = \ell \in [3,\sqrt{N}]\subseteq [\xi_{N,0},\xi_{N,n}]\equiv [\displaystyle\frac{N o_{min}}{x_{0}},\frac{N o_{min}}{x_{0}} \ c^n]  \nonumber
\end{eqnarray}
or 
\begin{eqnarray}\label{range2}
&\mbox{Method $(2)$:} \\ &\xi_N = \ell \in[\sqrt{N},N]\subseteq [\xi_{N,0},\xi_{N,n}] \equiv \displaystyle[\frac{N o_{min}}{x_{0}},\frac{N o_{min}}{x_{0}} \ c^n] \nonumber.
\end{eqnarray}

Let us consider first the method $(1)$. The lowest trial factor $3$ to be checked leads to the condition
\begin{eqnarray}\label{xiN01}
\xi_{N,0} \equiv \frac{N o_{min}}{x_{0}}\leq 3,
\end{eqnarray}
implying
\begin{eqnarray}\label{x0N1}
\frac{x_{0}}{o_{min}}\geq \frac{x_{0N}^{(1)}}{o_{min}}\equiv \frac{N}{3}.
\end{eqnarray}
In an analogous way, the value $\sqrt{N}$ of the largest trial factor  to be checked in Eq. (\ref{range1}) determines the condition
\begin{eqnarray}\label{xiNn1}
\xi_{N,n} \equiv \frac{N o_{min}}{x_{0}} \ c^n \geq \sqrt{N}.
\end{eqnarray}
This, together with  Eq. (\ref{x0N1}), implies the minimum number 
\begin{eqnarray}\label{nN1}
n^{(1)}_{N,x_0} &\equiv &\ceil*{log_{c}\frac{x_{0}}{ o_{min}\sqrt{N}}}\nonumber \\
&\leq &\ceil*{log_{c}\frac{\sqrt{N}}{3}}\equiv n_{N,min}^{(1)}
\end{eqnarray} 
 of necessary  interferograms $I^{(M,j)}(o_{\xi};x_i)$, with $i=0,1,...,n-1$,   to factor a generic integer $N$  with the method (1).

We consider now the  method $(2)$ in Eq. (\ref{range2}).
The lowest trial factor $\sqrt{N}$ to be checked leads to the condition
\begin{eqnarray}\label{xiN02}
\xi_{N,0} \equiv \frac{N o_{min}}{x_{0}}\leq \sqrt{N},
\end{eqnarray}
which implies
\begin{eqnarray}\label{x0N2}
\frac{x_{0}}{o_{min}}\geq \frac{x_{0N}^{(2)}}{o_{min}}\equiv  \sqrt{N}.
\end{eqnarray}
In an analogous way, the value $N$ of the largest trial factor  to be checked in Eq. (\ref{range2}) determines the condition
\begin{eqnarray}\label{xiNn2}
\xi_{N,n}\equiv \frac{N o_{min}}{x_{0}} \ c^n \geq N.
\end{eqnarray}
This, together with Eq. (\ref{x0N2}), leads to the minimum number 
\begin{eqnarray}\label{n2}
n^{(2)}_{x_0} &\equiv &\ceil*{log_{c}\frac{x_{0}}{o_{min}}}
\geq  \ceil*{log_{c} \sqrt{N}}\equiv n_{N,min}^{(2)}
\end{eqnarray} 
 of necessary  interferograms $I^{(M,j)}(o_{\xi};x_i)$, with $i=0,1,...,n-1$,   to factor a generic integer $N$ with the method (2).

This demonstrates that, in principle, the number $n$ of interferograms necessary for factorizing an arbitrary large number $N$, using a given range  $[o_{min},o_{max}]$ of the physical variable $o_{\xi}$, scales logarithmically with respect to  $\sqrt{N}$ and thereby polynomially with respect to the number of binary digits associated with $N$. It is important to point out that the narrower the range $o_{min}\leq o_{\xi}\leq o_{max}$ is, the closer the value of $c$ in Eq. (\ref{c2}) is to $1$ and thereby the larger is the value of $n$ in Eq. (\ref{n}). However, it would be enough to cover a spectrum such that $o_{max}\equiv2 o_{min}$  in order to obtain a scaling that is logarithmic (base $2$) and thereby polynomial with respect to the number of binary digits associated with $N$.

\subsubsection{Extension to the factorization of an exponential number of integers}

So far we have considered the case of factoring a single number with a sequence of $n$ ``factoring'' interferograms. However, the scaling property $\xi_N\equiv N o_{\xi}/x$ implies, as pointed out before, that we can exploit the same interferograms in order to factor not only a single integer but any integer in any given range $N_{min}\leq N \leq N_{max}$. We determine how, in such a case, the number $n$ of experimental runs depends on the smallest and the largest number $N_{min}$ and $N_{max}$ that can be factored.

In particular, factorization can be achieved for all  values 
of  $N$, with $N_{min}\leq N \leq N_{max}$,  only if for each single value is satisfied either the condition   (\ref{range1}) for the method $(1)$ or the condition  (\ref{range2}) for the method $(2)$.
Let us consider first the method $(1)$ in Eq. (\ref{range1}). The condition in Eq. (\ref{xiN01}) needs to be satisfied
for each $N_{min}\leq N \leq N_{max}$, which implies
$$ \frac{N_{max} o_{min}}{x_{0}}\leq 3$$
leading to
\begin{eqnarray}\label{x01}
\frac{x_{0}}{o_{min}} \geq \frac{x_{0}^{(1)}}{o_{min}} \equiv \frac{N_{max}}{3}.
\end{eqnarray}
In an analogous way, the condition in Eq. (\ref{xiNn1}) needs to hold
for each $N_{min}\leq N \leq N_{max}$, leading to
$$  \frac{ o_{min}}{x_{0}} \ c^n \geq \frac{1}{\sqrt{N_{min}}}.$$
This, together with Eq. (\ref{x01}), implies the minimum number
\begin{eqnarray}\label{nx01}
n_{x_0}^{(1)} &\equiv & \ceil* {log_{c} \ \frac{x_{0}}{o_{min}\sqrt{N_{min}}}} \nonumber\\
&\geq &\ceil*{log_{c}\ \frac{N_{max}}{3\sqrt{N_{min}}}}\equiv n^{(1)}_{min}
\end{eqnarray}
 of necessary  interferograms $I^{(M,j)}(o_{\xi};x_i)$, with $i=0,1,...,n-1$,   to factor all the integers $N$ in any given interval  $N_{min}\leq N \leq N_{max}$ with the method (1). 

We consider now the method $(2)$ in Eq. (\ref{range2}).
The condition (\ref{xiN02}) needs to be satisfied 
for each $N_{min}\leq N \leq N_{max}$, which implies
$$ \frac{ o_{min}}{x_{0}}\leq \frac{1}{\sqrt{N_{max}}}$$
and thereby 
\begin{eqnarray}\label{x02}
\frac{x_{0}}{o_{min}} \geq \frac{x_{0}^{(2)}}{o_{min}}\equiv \sqrt{N_{max}}.
\end{eqnarray}
In an equivalent way, Eq. (\ref{n2}) reads
\begin{eqnarray}\label{n2}
n^{(2)}_{x_0} &\equiv &\ceil*{log_{c}\frac{x_{0}}{o_{min}}}
\geq  \ceil*{log_{c} \sqrt{N_{max}}}\equiv n_{min}^{(2)}
\end{eqnarray} 
of necessary  interferograms $I^{(M,j)}(o_{\xi};x_i)$, with $i=0,1,...,n-1$,   to factor all the integers $N$ in any given interval  $N_{min}\leq N \leq N_{max}$ with the method (2).

In conclusion, the described algorithm allows in both method (1) and (2) the factorization of an exponential number
$$\Delta N \equiv N_{max} - N_{min} \sim N_{max}\sim 2^{n_{max}}$$
 of integers, with $n_{max}$ number of binary digits of $N_{max}$, by using a polynomial number of interferograms in a given physical domain $[o_{min},o_{max}]$. The largest factorable number is upper limited by the condition for $x_0/o_{min}$ in Eq. (\ref{x01}) and Eq. (\ref{x02}).

\subsubsection{Example of factorization of  $N \leq N_{max}\equiv 64$}

We now describe the implementation of our factoring algorithm  for  $N_{max}\equiv 64$. We consider a generic observable $O_{\xi}$ with values $o_{\xi}=\xi x$  in a given range $o_{min} \leq o_{\xi} \leq o_{max}$ which satisfies the condition
\begin{eqnarray}\label{cexample}
c \equiv o_{max}/o_{min}=2.
\end{eqnarray}

We consider first the method $(1)$ in Eq. (\ref{range1}). It turns out that any integer in the range $N_{min}\equiv 8\leq N \leq N_{max}\equiv 64$  can be factored, according to Eq. (\ref{nx01}), by exploiting
$$n=n^{(1)}_{min} \equiv \ceil*{log_{2} \frac{64}{3\sqrt{8}}}=3$$
interferograms $I(o_{\xi};x_{i}^{(1)})\equiv I^{(M=3,j=2)}(o_{\xi};x_{i}^{(1)})$ defined by Eq. (\ref{cespatternoxoxi}), with $i=1,2,3$, where the value $x=x_{0}^{(1)}$ associated with the first interferogram satisfies the condition in Eq. (\ref{x01})
\begin{eqnarray}\label{x01example}
\frac{x_{0}^{(1)}}{o_{min}} \equiv \frac{N_{max}}{3}=\frac{64}{3}.
\end{eqnarray}
 From Eq. (\ref{xiN}) we can now determine all the values
\begin{eqnarray}\label{xi1example}
 x=x_{i}^{(1)}\equiv c^{-i} x_{0}^{(1)} = \frac{64}{3} \cdot 2^{-i}  o_{min},
\end{eqnarray}
with $i=0,1,2$, for each of the $n=3$ interferograms. 
We assume that our physical system is able to record such interferograms in the range $o_{min} \leq o_{\xi} \leq 2 o_{min}$. In Fig. $5$  we simulate these interferograms, which are able to cover all the trial factors in the interval $[3, \sqrt{N}]$ for any integer in the range $N_{min}\equiv 8\leq N \leq N_{max}\equiv 64$. An example is given for the factorization of $N=15,63$ by rescaling the axis $o_{\xi}$ as a function of $\xi_{15}$ and $\xi_{63}$ according to Eq. (\ref{xiN}). The trial factors are marked, respectively, with continuous lines and dashed lines. We find the factors $3$ and $5$ of $15$ corresponding to maxima of the interferogram $I^{(M,j)}(o_{\xi};x_2^{(1)})$. On the other hand the factor $3$ of $63$ is associated with the maximum of the interferogram $I^{(M,j)}(o_{\xi};x_0^{(1)})$,
while the factors $7$ and $9$ emerge from the maxima in the interferogram $I^{(M,j)}(o_{\xi};x_1^{(1)})$.

We consider now the method $(2)$ in Eq. (\ref{range2}). According to Eq. (\ref{n2}), any integer in the range $N_{min}\equiv 1\leq N \leq N_{max}\equiv 64$ can be factored by exploiting the same number 
$$n = n^{(2)}_{min} \equiv \ceil*{log_{2} \sqrt{64}}=3$$
of interferograms  $I(o_{\xi};x_{i}^{(2)})\equiv I^{(M=3,j=2)}(o_{\xi};x_{i}^{(2)})$ defined by Eq. (\ref{cespatternoxoxi}), with $i=1,2,3$, where the value $x=x_{0}^{(2)}$ associated with the first interferogram satisfies the condition in Eq. (\ref{x02})
\begin{eqnarray}\label{x02example}
\frac{x_{0}^{(2)}}{o_{min}} \equiv \sqrt{N_{max}} = 8.
\end{eqnarray}
 From Eq. (\ref{xiN}) we can now determine all the values
\begin{eqnarray}\label{xi2example}
 x=x_{i}^{(1)}\equiv c^{-i} x_{0}^{(1)} =8\cdot 2^{-i} o_{min},
\end{eqnarray}
with $i=0,1,2$, for each of the $n=3$ interferograms. 
We assume that our physical system covers the spectrum $o_{min} \leq o_{\xi} \leq 2 o_{min}$ of values of a given observable $O_{\xi}$. In Fig. $6$  we simulate such interferograms, which are able to cover all the trial factors in the interval $[ \sqrt{N},N]$ for any integer in the range $N_{min}\equiv 1\leq N \leq N_{max}\equiv 64$. We again give an example for the factorization of $N=15,63$ by rescaling the axis $o_{\xi}$ as a function of $\xi_{15}$ and $\xi_{63}$ according to Eq. (\ref{xiN}). We find that the factor $3$ of $15$ corresponds to a maximum of the interferogram $I^{(M,j)}(o_{\xi};x_0^{(2)})$, while the factor $5$ is associated with a maximum of the interferogram $I^{(M,j)}(o_{\xi};x_1^{(2)})$. On the other hand the factor $9$ of $63$ is associated with the maximum of the interferogram $I^{(M,j)}(o_{\xi};x_0^{(2)})$.

\begin{figure*}[h]
   \begin{center}
   \begin{tabular}{c}
  \includegraphics[width=1\textwidth,natwidth=610,natheight=642]{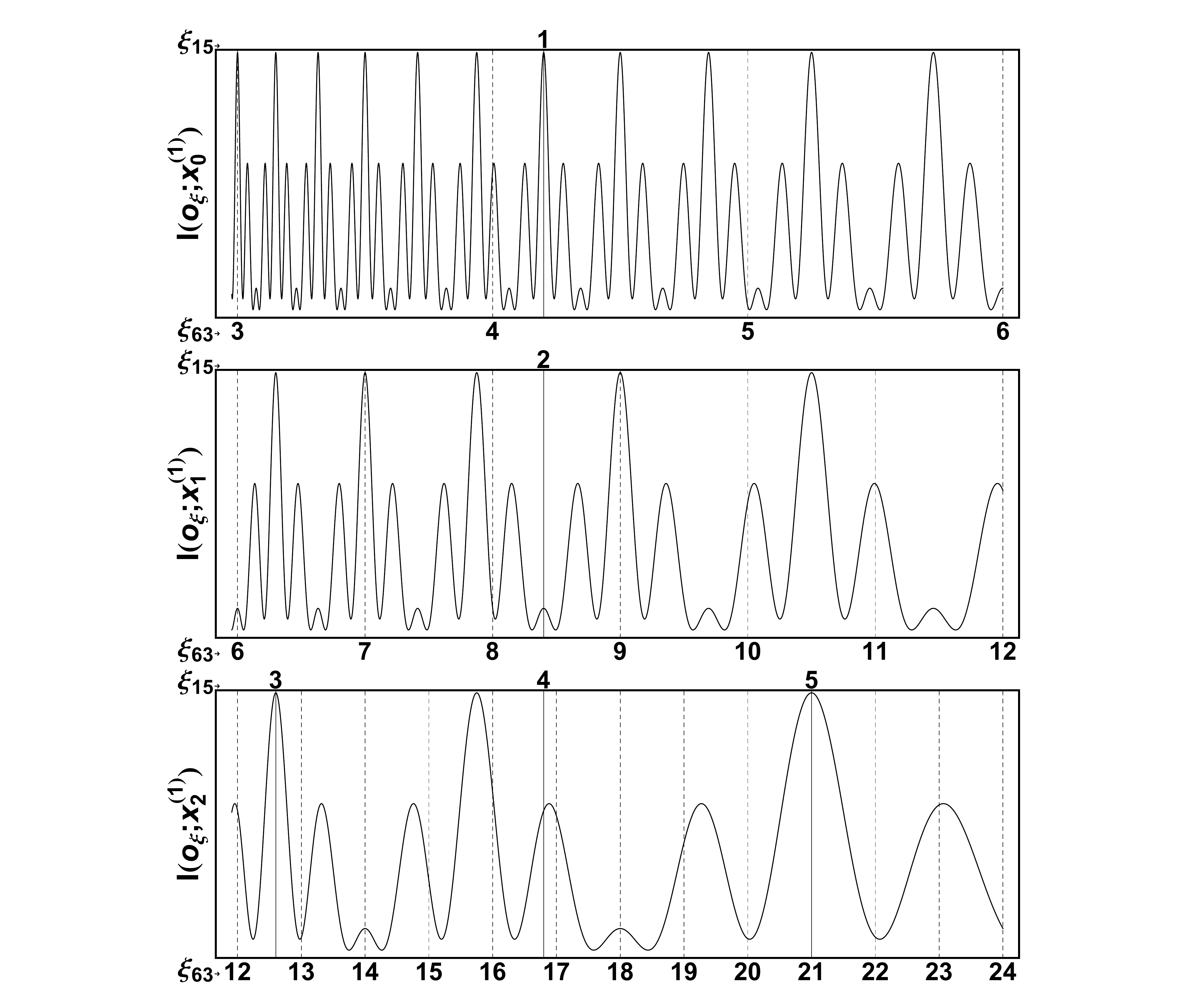}
   \end{tabular}
   \end{center}
\begin{quote}\caption  {Factorization of any integer in the range $N_{min}\equiv 8\leq N \leq N_{max}\equiv 64$ by using the method $(1)$ aimed at checking all the trial factors in the range $[3,\sqrt{N}]$. We exploit $n=3$ interferograms $I(o_{\xi};x_{i}^{(1)})\equiv I^{(M=3,j=2)}(o_{\xi};x_{i}^{(1)})$ defined by Eq. (\ref{cespatternoxoxi}), with $i=1,2,3$, where the range of values $o_{\xi}$ satisfies the condition (\ref{cexample}) and the values $x_{i}^{(1)}$ are given by Eq. (\ref{xi1example}).  We give an example for the factorization of $N=15,63$ by rescaling the axis $o_{\xi}$ as a function of $\xi_{15}$ and $\xi_{63}$ according to Eq. (\ref{xiN}). The trial factors correspond, respectively, to continuous lines and dashed lines. We find the factors $3$ and $5$ of $15$ corresponding to maxima of the interferogram $I^{(M,j)}(o_{\xi};x_2^{(1)})$. On the other hand, the factor $3$ of $63$ is associated with the maxima of the interferogram $I^{(M,j)}(o_{\xi};x_0^{(1)})$, while the factors $7$ and $9$ emerge from the maxima in the interferogram $I^{(M,j)}(o_{\xi};x_1^{(1)})$.\label{example1}}
   \end{quote}\end{figure*}
\begin{figure*}[h]
   \begin{center}
   \begin{tabular}{c}
   \includegraphics[width=1\textwidth,natwidth=610,natheight=642]{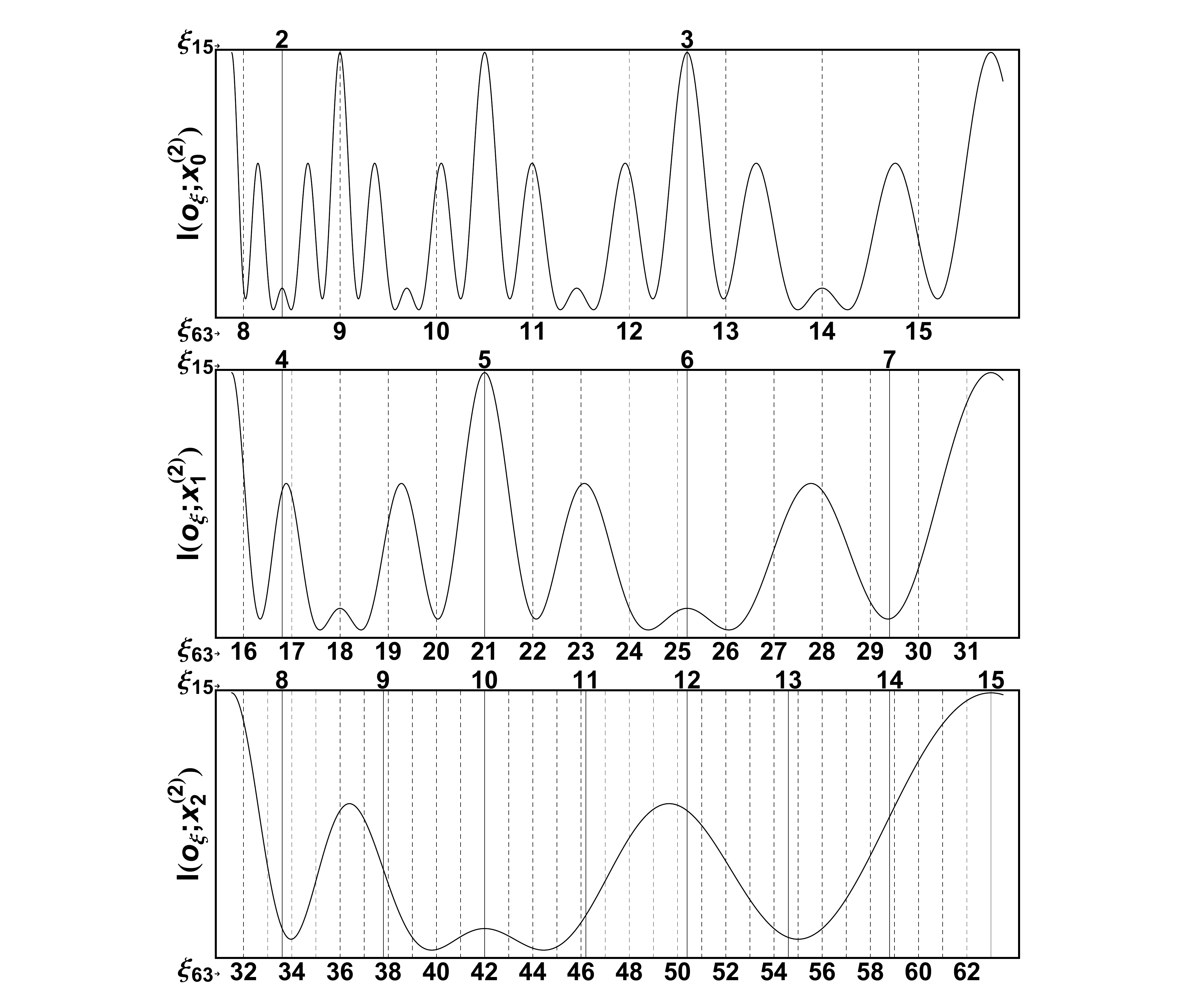}
   \end{tabular}
   \end{center}
\begin{quote}\caption  {Factorization of any integer in the range $N_{min}\equiv 1\leq N \leq N_{max}\equiv 64$ by using the method $(2)$  aimed at checking all the trial factors in the range $[\sqrt{N},N]$. We exploit $n=3$ interferograms $I(o_{\xi};x_{i}^{(2)})\equiv I^{(M=3,j=2)}(o_{\xi};x_{i}^{(2)})$ defined by Eq. (\ref{cespatternoxoxi}), with $i=1,2,3$, where the range of values $o_{\xi}$ satisfies the condition (\ref{cexample}) and the values $x_{i}^{(2)}$ are given by Eq. (\ref{xi2example}). We find that the factor $3$ of $15$ corresponds to a maximum of the interferogram $I^{(M,j)}(o_{\xi};x_0^{(2)})$, while the factor $5$ is associated with a maxima of the interferogram $I^{(M,j)}(o_{\xi};x_1^{(2)})$. On the other hand the factor $9$ of $63$ is associated with the maximum of the interferogram $I^{(M,j)}(o_{\xi};x_0^{(2)})$.\label{example2}}
   \end{quote}\end{figure*}

\section{Generalization of the CTES algorithm}\label{generalalgo}

In this section,  we will introduce a generalization of the CTES algorithm. 
In particular, we generalize the scaling property $\xi_N \equiv N \xi$ defining the new auxiliary variable
\begin{eqnarray}\label{scalingeneral}
\xi_{N,s}\equiv s \xi_N\equiv s  N \xi,
\end{eqnarray}
where $s$ is equal to the product of one or more generic prime numbers $p_k$ ($s\equiv\Pi_k p_k$).
We can now rescale the CTES function in Eq. (\ref{cespattern}) as a function of the continuous variable $\xi_{N,s}$:
\begin{eqnarray}\label{rescaledcespatterns}
{\cal{I}}^{(M,j)}(\xi_{N,s}) = |\frac{1}{M} \sum_{m=1}^{M} \exp\left[
2\pi  i (m-1)^j \frac{N s}{\xi_{N,s}}\right]|^2.
\end{eqnarray}

Let us describe how such a rescaled function allows us to find the factors of a generic number $N$. If one of the chosen prime numbers $p_k$ that defines the value of $s$ is a factor of $N$, we have solved the problem. If not, the factors of $N$ are the integer values of $\xi_{N,s}$, different from the prime numbers $p_k$ that correspond to dominant maxima in the rescaled CTES function given by Eq. (\ref{rescaledcespatterns}).

In a general analogue implementation, the auxiliary variable $\xi_{N,s}$ in Eq. (\ref{scalingeneral}) is obtained by rescaling the values $o_{\xi}\equiv \xi x$ of the observable $O_{\xi}$ according to
\begin{eqnarray}\label{scaling2oxi}
\xi_{N,s}\equiv  \frac{s N}{x} o_{\xi}.
\end{eqnarray}

Thereby, the range $[\xi_{N,s}^{(min)}, \xi_{N,s}^{(max)}]$ of trial factors covered by the variable $\xi_{N,s}$ can be equivalent to the interval $[\xi_{N}^{(min)}, \xi_{N}^{(max)}]$ of trial factors covered by the variable $\xi_{N}$ by exploiting a range for the variable $o_{\xi}$ of the physical observable $O_{\xi}$ of length $\Delta o = o_{max} - o_{min}$ reduced by a factor $s$. 
However, at the same time, each of the values $x=x_i$, with $i=0,1,...,n-1$, increases by the same factor $s$ according to the iteration formula (\ref{x_iN}) where $x_0$ in Eq. (\ref{x02}) now reads
\begin{eqnarray}\label{x02s}
x_{0}\equiv s  N_{max} o_{min}/\xi_{0},
\end{eqnarray}
with $\xi_0 \equiv 1, \sqrt{N_{min}}$ depending on the range (\ref{range1}) or (\ref{range2}), respectively, of trial factors we are considering. 
This implies that the maximum factorable integer $N_{max}$ is upper limited by $x_0/(s o_{min})$, where $x_{max}$ is the maximum value physically allowed for the parameter $x$.


\section{Remarks}

We have described a novel analogue  algorithm based on the experimental measurement of CTES interferograms $I^{(M,j)}(o_{\xi};x)$ depending on the continuous argument $o_{\xi}$ and the discrete parameter $x$ associated with two suitable physical observables $O_{\xi}$ and $O_x$, respectively. The domain of factorable integers is determined by the range $o_{min}\leq o_{\xi}\leq o_{max}$ of the continuous variable $o_{\xi}$ and by the maximum value $x_{max}$ allowed by the parameter $x$ .

The largest integer $N_{max}$  factorable with a single  CTES interferogram $I^{(M,j)}(o_{\xi};x)$ associated with a given value of $x$ and defined in a certain range $o_{min}\leq o_{\xi}\leq o_{max}$ is upper limited by the value $(o_{max}/o_{min})^2$. For larger integers $N$  the available physical spectrum of observable values $o_{\xi}$ may not be  enough to cover all the trail factors either in the range $1\leq \xi_N \leq \sqrt{N}$ or $\sqrt{N}\leq \xi_N \leq N$. For this reason, we have introduced an algorithm based on the measurement of $n$ different interferograms $I^{(M,j)}(o_{\xi};x_i)$, with $i=0,1,...,n-1$, associated with the respective values $x=x_i$. In this case the largest number factorable is upper limited by the value $x_{max}/o_{min}$.  We have demonstrated that the number $n$ of necessary experimental runs scales logarithmically with respect to the root of the number $N$ we want to factor.   Very interestingly our method allows a parallel factorization of an exponential number of large integers with respect to the number of binary digits $n_{max}$ associated with $N_{max}$.
These results are very important in view of the optical implementation of the algorithm described in Ref. \cite{pra2}.

Moreover, we have introduced a generalized CTES procedure defined by the scaling property $\xi_{N,s}\equiv s  N \xi$, where $s\equiv\Pi_k p_k$ with $p_k$ generic prime numbers which are non factors of $N$. We have shown that it is possible to reduce for a factor $s$ the length $\Delta o = o_{max} - o_{min}$ of the range of values $o_{\xi}$ in which the function $I^{(M,j)}(o_{\xi};x)$ is recorded. In such a case, the maximum factorable integer $N_{max}$ is upper limited by $x_0/(s o_{min})$.

In Ref. \cite{pra2}, we describe in detail how an optical computer enables a physical computation of the CTES algorithm for several orders $j$. In such a case  the values $o_{\xi}$ and $x$ defining the analogue interferogram $I^{(M,j)}(o_{\xi};x)$ correspond, respectively, to the wavelengths $\lambda$ of a polychromatic source and to the unit of displacement defining the optical paths in a generalized Michelson interferometer \cite{jmp}. Indeed, an experimental proof of the principle of the  CTES algorithm has been performed in the case of $j=2$ and $j=3$, leading to the factorization of seven-digit numbers \cite{pra2,prl}.

\section{Towards a polynomial scaling in the number of physical resources}\label{quantum}

We point out that any ``classical" implementation of the algorithm described so far would lack exponential speed-up.
Indeed, the largest number factorable $N_{max}$ is upper limited either by the value $(o_{max}/o_{min})^2$ or $x_0/o_{min}$, depending on the use of a single interferogram or a sequence of interferograms.

Moreover, the accuracy  in the variable $\xi$ in Eq. (\ref{oxi})
$$\Delta \xi = \frac{o_{\xi}}{x^2} \Delta x + \frac{1}{x} \Delta o_{\xi}\leq  \frac{ o_{max}}{x^2} \Delta x + \frac{1}{x} \Delta o_{\xi}$$
 depends on the given experimental indeterminations $\Delta o_{\xi}$ and $\Delta x$ associated with the measurement of the observables $O_{\xi}$ and $O_x$, respectively. This shows from Eqs. (\ref{deltaxiN1}) and (\ref{deltaxiN2}) that the unit $x$ defining the CTES interferograms in Eq. (\ref{cespatternoxoxi}) needs to grow exponentially with respect to the number of bits associated with the largest number $N_{max}$ in the interval of integers to be factored.   

The ability to resolve the maxima associated with factors from non factors for larger and larger integers $N$  \cite{thesis} can be, in general, improved by implementing "factoring" interferograms  of either a larger order $j$ or with a larger number $M$ of interfering terms as can be inferred by the relative curlicue functions in Fig. $1$.

Ultimately, only a ``quantum" system able to encode such physical observables in a qubit representation in line with Shor's algorithm would make it possible to avoid the requirement of exponentially large values for the observable $O_x$. 
In particular, a crucial role in our algorithm is played by the hyperbolic function $f(\xi)$ in Eq. (\ref{f(xi)}) emerging from the ratio of the values $o_x$ and $o_\xi$ and  characterizing the ``factoring" CTES interferograms in Eq. (\ref{cespatternoxoxi}).
A ``qubit" parallel representation of  functions analogous to the curlicue function may lead to novel factoring algorithms based on a  polynomial number of resources.  Moreover, multi-qubit quantum interference may serve as an efficient tool to distinguish factors from non factors. 
\begin{acknowledgements}
We thank  H. Zhang, X. He, Y.H. Shih and W. P. Schleich for their contributions on this topic and J. Franson, M. Freyberger, A. Garuccio, S. Lomonaco, R. Meyers, T. Pittman and M. H. Rubin for many fruitful discussions.
\end{acknowledgements}


\end{document}